\documentclass[conference]{IEEEtran}
\IEEEoverridecommandlockouts
\usepackage{cite}
\usepackage{amsmath,amssymb,amsfonts}
\usepackage{algorithmic}
\usepackage{graphicx}
\usepackage{textcomp}
\usepackage{enumitem}
\usepackage{verbatim}
\usepackage{multirow}
\usepackage{xcolor}
\usepackage{cuted}
\usepackage{mathrsfs}
\def\BibTeX{{\rm B\kern-.05em{\sc i\kern-.025em b}\kern-.08em
    T\kern-.1667em\lower.7ex\hbox{E}\kern-.125emX}}
\usepackage{etoolbox}
\makeatletter
\patchcmd{\thebibliography}
  {\section*{\refname}}
  {\section*{\refname}}
  {}{}
\makeatother
\begin{document}
\raggedbottom
\title{A Three-Stage Unsupervised Online Physics-Informed Neural Network for Microgrid Optimal Power Flow \\
}
\author{\IEEEauthorblockN{Abinash Giri}
\IEEEauthorblockA{\textit{Electrical and Computer Engineering} \\
\textit{Michigan Technological University}\\
Houghton, MI 49931 \\
akgiri@mtu.edu}
\and
\IEEEauthorblockN{Yi Hu}
\IEEEauthorblockA{\textit{Electrical and Computer Engineering} \\
\textit{Michigan Technological University}\\
Houghton, MI 49931 \\
yhu6@mtu.edu}
}
\maketitle
\begin{abstract}
   Optimal allocation of generators is a key challenge in power systems to achieve economical dispatch while satisfying operational constraints. Conventional optimization techniques can be computationally expensive, limiting their applicability in real-time environments, and neural networks often require offline retraining under new operating conditions. This paper proposes a three-stage unsupervised online physics-informed neural network (PINN) framework for optimal power flow (OPF). Simulation results demonstrate that the proposed framework satisfies load demand and system constraints while achieving a $\$43,665$ lower operating cost than \texttt{fmincon}-based optimization under the same operating conditions. The training process analysis further demonstrate the framework's ability to adapt its parameters online to variations in load demand and renewable generation in real time.
\end{abstract}
\begin{IEEEkeywords}
microgrid, Newton-Raphson, online unsupervised learning, optimal power flow, physics-informed neural network.
\end{IEEEkeywords}
\section{Introduction}
\label{sec:introduction}
The increasing penetration of distributed energy resources (DERs) provides an opportunity for independent system operators (ISOs) to reduce dependence on conventional generation and potentially lower system operating costs. However, different operating characteristics and cost functions of different generators creates challenges to achieve economic resource allocation\cite{ref1}. Optimal Power Flow (OPF) is widely used to determine an economically efficient operating point while satisfying network and operational constraints\cite{ref2}.

Alternating Current Optimal Power Flow (AC-OPF) is a nonlinear and nonconvex optimization problem whose computational burden increases with the number of generators, decision variables, and operational constraints \cite{ref3}. In addition, repeatedly solving AC-OPF under changing load and generation conditions is computationally expensive, limiting its suitability for real-time power system operation. To address this challenge, machine learning-based approaches, particularly neural networks (NNs), have been investigated for OPF applications. Although NN-based approaches can provide fast inference, they require representative training data, which is difficult or expensive to obtain. Moreover, conventional NNs do not enforce the physical and operational constraints of the power system and produce infeasible operating points \cite{ref4}.

Physics-Informed Neural Networks (PINNs) provide a promising alternative by incorporating physical information into the NN learning process \cite{ref5}. Power system constraints can be embedded into the training objective, allowing the network to learn from the underlying physical relationships of the power system while reducing its dependence on large labeled datasets. However, those methods still require large amount of well labeled data, and the performance of an offline-trained network degrades when the operating conditions differ from those represented in the training data.

Several studies have investigated physics-informed and unsupervised learning approaches for AC-OPF. In \cite{ref6}, an unsupervised online PINN framework was proposed using two interconnected neural networks, with one serving as the AC-OPF solver and the other as the AC power flow (ACPF) evaluator. In \cite{ref7}, a graph-based neural network framework was proposed for real time AC-OPF, where the network predicts generator active power and voltage setpoints and uses online fine-tuning to adapt to time varying operating conditions. However, both frameworks require an offline pre-training before deployment. Furthermore, both approaches report higher generation costs when compared to conventional optimization benchmarks.

\begin{figure*}[!t]
    \centering
    \includegraphics[width = 0.85\textwidth]{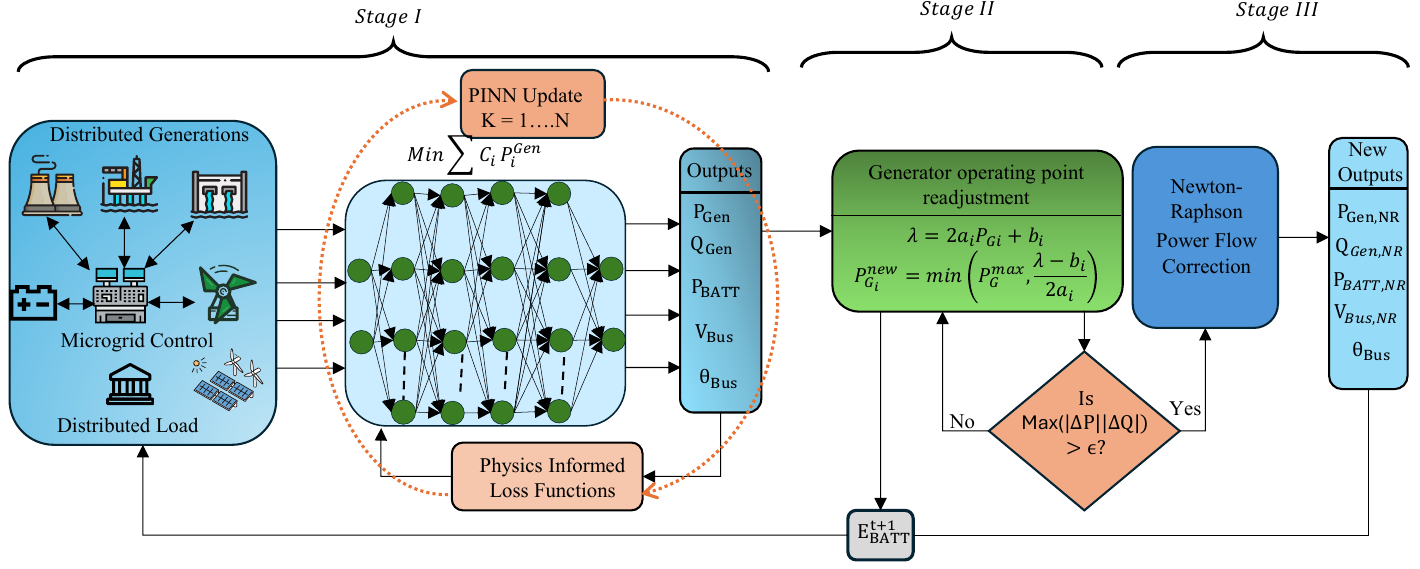}
    \caption{Proposed three stage framework combining online PINN, economic redispatch, and a conditional NR correction for microgrid OPF.}
    \vspace{-6pt}
    \label{fig:proposed_framework}
\end{figure*}

To address these limitations, this paper proposes a three-stage unsupervised online PINN framework, which does not require pre-training before deployment, for microgrid OPF. The framework combines online physics-informed learning, marginal-cost-based economic redispatch, and a conditional Newton-Raphson (NR) correction to improve economic operation and AC power flow feasibility.
The main contributions of this paper are:
\begin{itemize}[nosep, leftmargin=1.2em]
    \item \textbf{An three-stage unsupervised PINN framework for AC-OPF} is proposed. The framework incorporates a PINN to set initial operation point, a marginal-cost-based redispatch to adjust generator setpoints, and a conditional NR correction to restore AC power flow feasibility.
    \item \textbf{An online training strategy} is designed to process sequential system data without labeled OPF solutions or retraining when scenarios change. Multiple training iterations at each time step promote rapid convergence.
    \item \textbf{A physics-informed loss function} is developed to account for power flow equations, generation costs, and operation limits, guiding the model to cost minimization without physical constraint violations.
\end{itemize}
The remainder of this paper is organized as follows.
Section~\ref{sec:Method} presents the methodology.
Section~\ref{sec:Result} presents the simulation results. Finally, Section~\ref{sec:Conclusion} concludes the paper.

\section{Methodology}
\label{sec:Method}

This section presents the proposed three-stage approach for solving the OPF problem. The problem is formulated to minimize the total operating cost of the generators and battery subject to the system operating constraints. The objective function is expressed as
\begin{flalign}
\min \sum_{t\in T}
\left(
a_g P_g^2(t)
+b_g P_g(t)
+c_g
\right)
+
K_B P_B(t),
\label{eq:objective}
\end{flalign}
where $a_g$, $b_g$, and $c_g$ are the quadratic, linear, and constant cost coefficients of generator $g$, respectively, and $K_B$ represents the battery cycling cost. 

As shown in Fig. \ref{fig:proposed_framework}, the proposed framework consists of three stages. In the first stage, a PINN takes sequential input data $x = \left[P_{L},Q_{L}, P_{PV}^{avail}, P_{wind}^{avail}, SOC\right]^T$and determines initial OPF solutions. The PINN is trained from scratch in an online manner. In the second stage, the active power setpoints of the conventional generators are readjusted based on their marginal costs to reduce the overall cost. In the third stage, a NR correction is applied when the post-dispatch mismatch calculated from allocated generation, load and calculated power injections exceeds a predefined tolerance to further improve the feasibility.

Battery SOC is updated at each time step based on the remaining battery energy and output from the previous step:
\begin{flalign}
    SOC = \frac{E_{t+1}}{E_{max}},
\end{flalign}

\begin{flalign}
    \begin{aligned}
        E_{t+1} &= E_{t} +\left(\eta_{ch} P_{B,t}^{ch} - \frac{P_{B,t}^{dis}}{\eta_{dis}}\right)\Delta t
    \end{aligned},
\end{flalign}
where $E_{max}$ is the maximum capacity of the battery and $\eta$ is the battery efficiency.

\subsection{Stage I -- PINN-Based Economic Dispatch}\label{sec:PINN}
\subsubsection{Model Architecture}
A fully connected feedforward neural network with four hidden layers containing 128, 256, 512, and 256 neurons is employed. The input and output layers contain 27 and 50 neurons, respectively. ReLU is used as the activation function, and the learning rate is set to $1\times10^{-4}$. The network receives sequential inputs $x$ and predicts the active and reactive power dispatch of the generators, battery, bus voltage magnitudes, and phase angles as
\begin{flalign}
\hat{\mathbf{y}}_{t}^{(k)}
= f_{\boldsymbol{\theta}_{t}^{(k)}}(\mathbf{x}_{t}),
\label{eq:Prediction}
\end{flalign}
where $k$ denotes the index of training iteration within one time step. 

\subsubsection{Online Training Process}
At each time step, the network parameters are updated through backpropagation using the Adam optimizer
\begin{flalign}
\boldsymbol{\theta}_{t}^{(k+1)}
= \operatorname{Adam}\left(
\boldsymbol{\theta}_{t}^{(k)},
\nabla_{\boldsymbol{\theta}}\mathcal{L}_{t,\mathrm{total}}^{(k)}
\right).
 \label{eq:back_propagation}
\end{flalign}

\noindent The updated parameters are then used with the same input to generate a new prediction, allowing the network to further minimize the physics-informed loss. After completing $K$ iterations, the final prediction is obtained as
\begin{flalign}
 \hat{\mathbf{y}}_{t}^{(k+1)}
= f_{\boldsymbol{\theta}_{t}^{(k+1)}}(\mathbf{x}_{t}),
\qquad
\hat{\mathbf{y}}_{t}^{*}
= f_{\boldsymbol{\theta}_{t}^{(K)}}(\mathbf{x}_{t}).
 \label{eq:updated_final_prediction}
\end{flalign}
To improve computational efficiency, an adaptive number of iterations is assigned based on the relative variation between consecutive input vectors:
\begin{flalign}
 r_t =
\frac{\left\|\mathbf{x}_t-\mathbf{x}_{t-1}\right\|}
{\left\|\mathbf{x}_{t-1}\right\|+\epsilon},
 \label{eq:relative input variatoin}
\end{flalign}
where $\mathbf{x}_t$ and $\mathbf{x}_{t-1}$ represent the current and previous input vectors, respectively, and $\epsilon$ is a small positive value used to prevent division by zero. The updated network parameters are retained for the next incoming sample, and the identical online learning procedure is repeated sequentially.

\subsubsection{Loss Function Design}
The physics-informed loss function consisting of power flow, operating-cost, ramp-rate, battery, line-loss, power-balance, and Karush--Kuhn--Tucker (KKT) terms\cite{ref8}:
\begin{flalign}
&\begin{aligned}
\mathcal{L}_{\mathrm{total}}
&= \lambda_{PF}\mathcal{L}_{PF}
+ \lambda_{\mathrm{cost}}\mathcal{L}_{\mathrm{cost}}
+ \lambda_{\mathrm{ramp}}\mathcal{L}_{\mathrm{ramp}}
+ \lambda_{\mathrm{batt}}\mathcal{L}_{\mathrm{batt}}\\
&\quad + \lambda_{\mathrm{line}}\mathcal{L}_{\mathrm{line}}
+ \lambda_{\mathrm{bal}}\mathcal{L}_{\mathrm{bal}}
+ \lambda_{\mathrm{KKT}}\mathcal{L}_{\mathrm{KKT}}.
\end{aligned}
&& \label{eq:total_loss}
\end{flalign}

Power flow Loss ($\mathcal{L}_{PF}$) minimizes the active and reactive power mismatches between the dispatched bus power injections and those calculated from the AC power flow equations:
\begin{flalign}
\begin{aligned}
\mathcal{L}_{PF}
&=
\operatorname{MSE}(\Delta P,0)
+\operatorname{MSE}(\Delta Q,0) \\
&\quad
+\left(\max_i |\Delta P_i|\right)^2
+\left(\max_i |\Delta Q_i|\right)^2,
\end{aligned}
\end{flalign}
where, $(\Delta P = P_G - P_L - P_{AC})$ and $(\Delta Q = Q_G - Q_L - Q_{AC})$, $P_G, Q_G$ are the allocated active and reactive power generation, $P_L, Q_L$ are the active and reactive load demand, and $P_{AC}, Q_{AC}$ are the active and reactive powers injection to each bus calculated from the power flow using the voltage magnitudes and angles from the PINN.

Cost Loss ($\mathcal{L}_{cost}$) minimizes the operating cost of the generators and battery operation:
    \begin{flalign}
        \begin{aligned}
            \mathcal{L}_{cost} &= \frac{1}{1000}\sum_{t\in T}
            \left[\left(
            a_g P_g^2(t)
            +b_g P_g(t)
            +c_g
            \right)
            +
            K_B P_B(t)\right]\Delta{t}.
        \end{aligned}
    \end{flalign}

Ramp Loss ($\mathcal{L}_{ramp}$) penalizes changes in generator outputs that exceed the specified ramp limits between consecutive time steps:
    \begin{flalign}
        \begin{aligned}
            \mathcal{L}_{ramp} &=
            \sum_{g} \left[\frac{\operatorname{max}\left(0, |P_{g}^t -P_{g}^{t-1}|-R_{g}\right)}{R_{g} + \epsilon}\right]^2,
        \end{aligned}
    \end{flalign}
where $R_g$ is the ramp limit of the generators.

Battery Loss ($\mathcal{L}_{batt}$) enforces the specified operating limits on the battery SOC:
    \begin{flalign}
        \begin{aligned}
            \mathcal{L}_{batt} & = \left[\frac{\operatorname{max}(0, E_{min} -E_{t+1})}{E_{max} +\epsilon}\right]^2 \\ + & \quad  \left[\frac{\operatorname{max}(0, E_{t+1} -E_{max})}{E_{max} +\epsilon}\right]^2 .
        \end{aligned}
    \end{flalign}

Line Loss ($\mathcal{L}_{line}$) reduces active power losses across transmission lines:
\begin{flalign}
\begin{aligned}
\mathcal{L}_{line}
&=\frac{
\displaystyle\sum_{}\max\left(0,P_{ij}+P_{ji}\right)
}{
S_{base}+\epsilon
}.
\end{aligned}
\end{flalign}

Balance Loss ($\mathcal{L}_{bal}$) enforces the active power balance between generation, load demand, and line losses:
\begin{flalign}
\mathcal{L}_{bal}
=
\left[
\frac{\sum_i P_G^i-\sum_i P_L^i-\sum P_{loss}^{line}}
{S_{base}}
\right]^2.
\end{flalign}

KKT Loss ($\mathcal{L}_{KKT}$) promotes economic generation allocation by enforcing equal marginal costs for generators operating within their limits.

\begin{flalign}
\mathcal{L}_{KKT}
=
\frac{1}{N_g}
\sum_{g=1}^{N_g}
\left[
\left(\frac{r_g^{int}}{10}\right)^2
+
\left(\frac{r_g^{min}}{10}\right)^2
+
\left(\frac{r_g^{max}}{10}\right)^2
\right],
\end{flalign}
\begin{flalign}
\begin{aligned}
r_g^{int} &= MC_g-\lambda,
r_g^{min}=\max(0,\lambda-MC_g),\\
r_g^{max} &= \max(0,MC_g-\lambda),
\end{aligned}
&
\end{flalign}
where $MC_{g}$ is the marginal cost of the generators.

\subsection{Stage II : Equal Incremental Cost-Based Dispatch}
Stage 1 provides the generators operating points, but it does not guarantee the optimal allocation of the convention generators. In this paper, these outputs predicted by the PINN are reallocated using the incremental cost principle to improve the economic dispatch.
According to the Lagrangian method, the economic dispatch problem can be formulated as given below\cite{ref9}:
\begin{flalign}
 \mathscr{L} =
\sum_{g\in\mathcal{G}} C_g(P_g)
+\lambda\left(P_{\mathrm{load}}-\sum_{g\in\mathcal{G}}P_g\right).
 \label{eq:lagrangian}
\end{flalign}

\noindent The first-order optimality condition for generators operating within their limits is given as:
\begin{flalign}
     \frac{\partial\mathscr{L}}{\partial P_g}
    =\frac{\partial C_g}{\partial P_g}-\lambda=0,
    \qquad
    \lambda=\frac{\partial C_g}{\partial P_g}
    =2a_gP_g+b_g.
\label{eq:incremental_cost}
\end{flalign}

\noindent The new generation setpoints are calculated as:
\begin{flalign}
 P_g^* =
\operatorname{clip}\left(
\frac{\lambda-b_g}{2a_g},
P_g^{\mathrm{low}},P_g^{\mathrm{up}}
\right).
 \label{eq:dispatch_refinement}
\end{flalign}
With the updated setpoints, the total generator operating cost is minimized when the equal incremental cost condition is satisfied.
\subsection{Stage III: Conditional NR-Based Power Flow Correction}

Stage II modifies the active power dispatch of the conventional generators to reduce operating cost. Since the redispatch changes the generator active power outputs, the redispatched  operating point becomes inconsistent with the voltage magnitudes and phase angles predicted by the PINN \cite{ref10}. The active and reactive power mismatches are calculated from the total generation, total load, and the nodal power injection from AC power flow. If the maximum mismatch exceeds the prescribed tolerance, NR stage will activated.

The NR stage uses the redispatched generator outputs and the PINN-predicted voltage magnitudes and phase angles as the initial operating point. The Jacobian matrix is then used to update the voltage magnitudes and phase angles:

\begin{flalign}
\boldsymbol{\theta}^{(k+1)}
&=
\boldsymbol{\theta}^{(k)}
+
\Delta\boldsymbol{\theta}^{(k)},
\qquad
\mathbf{V}^{(k+1)}
=
\mathbf{V}^{(k)}
+
\Delta\mathbf{V}^{(k)}.
\label{eq}
\end{flalign}

Using the updated voltage magnitudes and phase angles, the AC power flow is recalculated to obtain the new active and reactive power mismatches. This process is repeated until the maximum mismatch is reduced below the prescribed tolerance. The final corrected generator outputs are then used to calculate the total operating cost.

\section{Simulation Results}
\label{sec:Result}

\begin{figure}
    \centering
    \includegraphics[width=0.89\columnwidth]{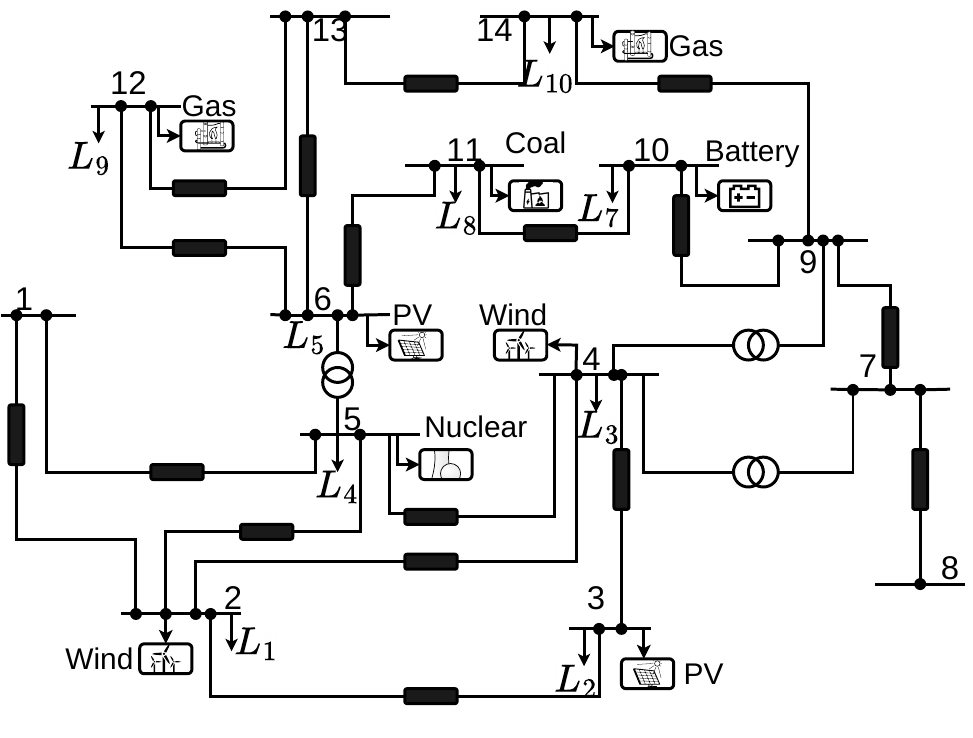}
    \vspace{-0.2cm}
    \caption{IEEE 14-bus system configuration used in this study.}
    \label{fig:IEEE Bus}
\end{figure}

The simulations are conducted using a modified IEEE 14-bus system operating in islanded mode, with the generator and load connections shown in Fig.~\ref{fig:IEEE Bus}. The load profile is scaled to a maximum of 85~MW. The capacities of the installed generators are 
$\left[P_{coal}, P_{NG},P_{URAN}, P_{PV}, P_{wind}, P_{B}\right]$ = $\left[10, 47, 28, 6 , 4.8, 2.5\right]$~MW.
All computations are performed on a 14-inch MacBook Pro running macOS Tahoe 26.6.2 with an Apple M4 Pro processor and 24 GB of unified memory. A conventional optimization method based on MATLAB's \texttt{fmincon} under the same operating conditions is selected as the baseline to evaluate the economic performance. This method solves the AC-OPF problem using the identical objective function and constraints as in proposed framework.  According to \cite{ref6, ref7}, existing physics-informed unsupervised learning-based approaches require an offline pre-training before deployment and report higher generation costs when compared to conventional optimization benchmarks, thus they are not adopted as baselines.

\subsection{Online Training Process Analysis}
This section illustrates how the total loss variates along the training process. During the cold-start stage, the first incoming sample was trained for a larger number of iterations to adapt the network to the initial operating condition. For subsequent samples, the number of training iterations was assigned according to the relative input variation defined in (\ref{eq:relative input variatoin}). Input changes below 1\% were assigned 5 iterations, changes between 1\% and 3\% were assigned 15 iterations, changes between 3\% and 5\% were assigned 25 iterations, and changes greater than or equal to 5\% were assigned 50 iterations.

\begin{figure}
    \includegraphics[width = \linewidth]{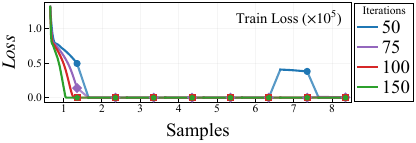}
    \vspace{-0.7cm}
    \caption{The training losses under different number of iteration configurations.}
\label{fig:Training Loss}
\end{figure}

To evaluate the network convergence under different numbers of cold-start iterations, the first sample was trained using 50, 75, 100, and 150 iterations. As shown in Fig.~\ref{fig:Training Loss}, the training loss for the 50 and 75 iteration cases converges within the second sample, indicating slower initial adaptation. Increasing the number of cold-start iterations to 100 reduces the convergence to the first sample. With 150 iterations, the loss reaches the lowest level within the first sample and remains stable for the subsequent samples. Based on this observation, 150 iterations were selected as the cold-start setting for the subsequent performance evaluation.

The computation times for 5, 10, 15, 25, and 150 iterations were 0.05, 0.09, 0.13, 0.23, and 1.23~s, respectively. These results demonstrate the fast computational performance of the proposed method and support its suitability for online implementation under changing operating conditions.

\begin{figure}[!t]
    \includegraphics[width = \linewidth]{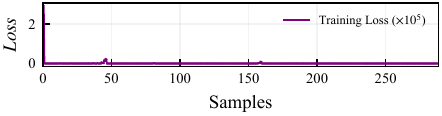}
    \vspace{-0.7cm}
    \caption{The training loss during complete one day simulation.}
\label{fig:Training Loss one day}
\end{figure}

\begin{figure}
    \includegraphics[width = \columnwidth]{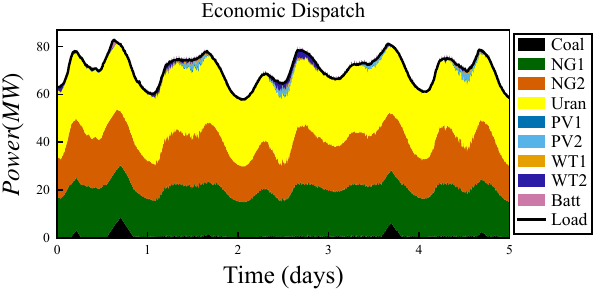}
    \caption{Economic dispatch of generation resources over five days.}
\label{fig:Economic Dispatch}
\end{figure}

The training loss over one day of sequential samples is given in Fig. ~\ref{fig:Training Loss one day}. The loss increased near the 45th sample. But the network reduced the loss by increasing the number of training iterations for that sample, after that the loss remained stable. This recovery demonstrates that the proposed online training framework can adapt to changes in operating conditions.

\subsection{OPF Results and Economic Performance Comparison}

\begin{figure}
    \includegraphics[width = \columnwidth]{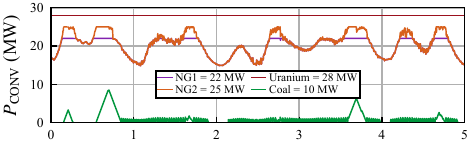}
    \includegraphics[width = \columnwidth]{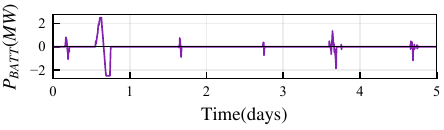}
    \caption{Conventional generators and battery active power outputs.}
    \label{fig:conventional power}
    \end{figure}

Fig.~\ref{fig:Economic Dispatch} shows that the aggregate dispatch generation follows the total load over the five-day window. Conventional generators supply most of the load, while wind and solar make smaller, variable contributions. As shown in Fig.~\ref{fig:conventional power}, the nuclear unit operates at its 28~MW limit throughout the period, while NG1 and NG2 vary their output and reach their respective limits of 22 and 25~MW during several intervals. The coal unit generates intermittently and remains below its 10~MW limit. Battery power stays near zero for most of the period, with brief charging and discharging events reaching approximately 2.5~MW in magnitude.



Table~\ref{tab:cost_comparison} compares cumulative operating costs between baseline and different versions of proposed model over 335 days to evaluate the contributions of each stage in the framework. The PINN alone reduces cost relative to FMINCON, and adding marginal-cost-based redispatch roughly doubles the savings. When NR correction is added (complete three-stage) the cost slightly increased when compared to PINN + Redispatch scenario, but the three-stage framework will guarantee the system in a feasible operating point. The operating point validation results will be shown in next section.


\begin{table}[!t]
\caption{Operating Cost Comparison Between Conventional Optimization Method and Proposed Method over 335 Days.}
\label{tab:cost_comparison}
\renewcommand{\arraystretch}{1}
\hspace*{\fill}%
\begin{tabular*}{\columnwidth}{@{\extracolsep{\fill}}|l|c|c|}
\hline
\textbf{Methods} & \textbf{Cost} & \textbf{Cost Reduction} \\
\hline
FMINCON (Baseline) & \$3,719,929 & N/A \\
\hline
PINN & \$3,698,201 & \$21,728 (0.584\%)\\
\hline
PINN+Redispatch &\$3,675,518  & \$44,411(1.194\%) \\
\hline
PINN+Redispatch+NR & \$3,676,272 & \$43,665 (1.174\%) \\
\hline
\end{tabular*}
\end{table}

\subsection{Operating Point Validation}

\begin{figure}
    \includegraphics[width = \columnwidth]{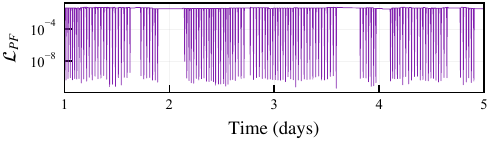}
    \vspace{-0.7cm}
    \caption{Power flow loss during online operation over five days.}
    \vspace{-6pt}
\label{fig:power_flow_loss}
\end{figure}

\begin{figure}
    \centering
\includegraphics[width=\columnwidth]{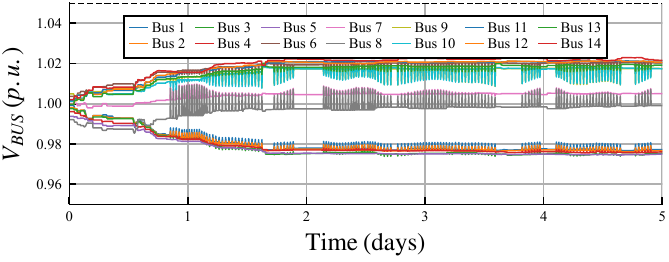}
    \caption{Voltage magnitudes at all 14 buses over the five-day window.}
    \vspace{-6pt}
    \label{fig:Bus Voltage}
\end{figure}



The dispatch result is validated for power flow loss, described in (9), and voltage violations to verify its feasibility.
Fig.~\ref{fig:power_flow_loss} shows that the power flow loss fluctuates during online operation, reaching the order of $10^{-2}$ at some samples and repeatedly dropping to the $[10^{-11}$, $10^{-10}]$ range. It show that the framework can optimize the power allocation for generators tha matches well with the loads during the five-day window. 
Fig.~\ref{fig:Bus Voltage} shows that the observed voltages range from approximately 0.97 to 1.02~p.u.. It indicates that there is no voltage violations at all 14 buses over the five-day window.


\section{Conclusion}\label{sec:Conclusion}
This paper presented a three-stage unsupervised online physics-informed neural network (PINN)-based framework for microgrid optimal power flow. The proposed framework maintains feasible operating conditions within the specified system constraints while improving the economic allocation of available generation resources. Under the same operating conditions, the proposed framework reduced the operating cost by $\$43,665$ when compared to a conventional optimization approach. The training-loss analysis show that the network adapts to changes in load demand and renewable generation while maintaining convergence. Future work will evaluate the scalability and investigate the performance of the proposed framework on large load dominated power systems.

\bibliographystyle{IEEEtran}
\bibliography{references/references}

@article{ref1,
  title={Data-driven optimal power flow: A physics-informed machine learning approach},
  author={Lei, Xingyu and Yang, Zhifang and Yu, Juan and Zhao, Junbo and Gao, Qian and Yu, Hongxin},
  journal={IEEE Transactions on Power Systems},
  volume={36},
  number={1},
  pages={346--354},
  year={2020},
  publisher={IEEE}
}

@article{ref2,
  title={A transparent neural network solver for optimal power flow problem},
  author={Hu, Zuntao and Zhang, Hongcai},
  journal={IEEE Transactions on Power Systems},
  volume={41},
  number={1},
  pages={54--69},
  year={2025},
  publisher={IEEE}
}

@article{ref3,
  title={Physics-Informed GNN for non-linear constrained optimization: PINCO a solver for the AC-optimal power flow},
  author={Varbella, Anna and Briens, Damien and Gjorgiev, Blazhe and D'Inverno, Giuseppe Alessio and Sansavini, Giovanni},
  journal={arXiv preprint arXiv:2410.04818},
  year={2024}
}

@article{ref4,
  title={Physics-informed neural networks for AC optimal power flow},
  author={Nellikkath, Rahul and Chatzivasileiadis, Spyros},
  journal={Electric Power Systems Research},
  volume={212},
  pages={108412},
  year={2022},
  publisher={Elsevier}
}

@article{ref5,
  title={Solving multi-period OPF with transmission interface constraints via physics-informed spatiotemporal graph convolutional networks},
  author={Rong, Hao and Qin, Zhijun},
  journal={International Journal of Electrical Power \& Energy Systems},
  volume={177},
  pages={111859},
  year={2026},
  publisher={Elsevier}
}

@article{ref6,
  title={Unsupervised Online Learning for AC Optimal Power Flow: A Gradient-Guided Physics-Informed Neural Network Approach},
  author={Jiang, Bozhen and Qu, Jing and Wang, Qin and others},
  year={2025},
  publisher={TechRxiv}
}

@article{ref7,
  title={UNION: A Unified AC-OPF Framework for Topology-Varying Real-Time Grid Operation},
  author={Park, Kyungnam and Song, Keunju and Lim, Yeji and Park, Suho and Kim, Kibaek and Kim, Hongseok},
  journal={arXiv preprint arXiv:2608.25784},
  year={2026}
}

@inproceedings{ref8,
  title={Optimal selection of generators in a microgrid for fuel usage minimization},
  author={Han, Yi and Young, Peter and Zimmerle, Daniel},
  booktitle={2013 IEEE Power \& Energy Society General Meeting},
  pages={1--5},
  year={2013},
  organization={IEEE}
}

@article{ref9,
    title={Cost-based droop schemes for economic dispatch in islanded microgrids},
  author={Chen, Feixiong and Chen, Minyou and Li, Qiang and Meng, Kaikai and Zheng, Yongwei and Guerrero, Josep M and Abbott, Derek},
  journal={IEEE Transactions on Smart Grid},
  volume={8},
  number={1},
  pages={63--74},
  year={2016},
  publisher={IEEE}
}

@article{ref10,
 title={Probabilistic deliverability assessment of distributed energy resources via scenario-based AC optimal power flow},
  author={Anton, Lauren{\c{t}}iu L and Ili{\'c}, Marija D},
  journal={Energies},
  volume={18},
  number={18},
  pages={4832},
  year={2025},
  publisher={MDPI}
}
\end{document}